\documentstyle[12pt,prl,aps]{revtex}
\newcommand{\be}{\begin{equation}}
\newcommand{\ee}{\end{equation}}
\newcommand{\bay}{\begin{eqnarray}}
\newcommand{\eay}{\end{eqnarray}}

\begin{document}
\begin{titlepage}
\begin{flushright}
IFUP-TH 45/96\\
May 1996\\
quant-ph/9605045
\end{flushright}

\vspace{5mm}

\begin{center}

{\Large \bf The tunneling time for a wave packet

\vspace{3mm}

as measured with a physical clock}

\vspace{12mm}
{\large Andrea Begliuomini$^{a}$ and Luciano Bracci$^{a,b}$\\}

\vspace{3mm}

$(a)$ Dipartimento di Fisica dell'Universit\`{a},\\
piazza Torricelli 2, I-56126 Pisa, Italy.\\
$(b)$ Istituto Nazionale di Fisica Nucleare, sezione di Pisa, Italy.\\
\end{center}

\vspace{2truecm}

\begin{quote}
\centerline{\bf Abstract}

We study the time required for a wave packet to tunnel beyond a square barrier, or to be reflected, by envisaging a physical clock which ticks only when the particle is within the barrier region. The clock consists in a magnetic moment initially aligned with the $x$ axis which in the barrier region precesses around a constant magnetic field aligned with the $z$ axis, the motion being in the $y$ direction. The values of the $x$ and $y$ components of the magnetic moment beyond or in front of the barrier allow to assign a tunneling or reflection time to every fraction of the packet which emerges from the barrier and to calculate tunneling times $\tau_{\rm T,x}$ and $\tau_{\rm T,y}$ and reflection times $\tau_{\rm R,x}$ and $\tau_{\rm R,y}$. The times $\tau_{\rm T,x}$ and $\tau_{\rm T,y}$ ($\tau_{\rm R,x}$ and $\tau_{\rm R,y}$) are remarkably equal, and independent of the initial position (in front of the barrier) of the packet.
\end{quote}

\vspace{0.5truecm}

\end{titlepage}
\clearpage

\section{Introduction}

An old problem in Quantum Mechanics which has attracted the attention of several authors (see \cite{hast89}-\cite{olre92}-\cite{lama94} for extensive reviews) is that of a satisfactory definition of the time required for such a peculiarly quantum mechanical phenomenon as tunnelling.
In a previous paper \cite{bebr96} we investigated the behaviour of a wave packet impinging onto a square barrier and proposed definitions of the interaction time (or dwell time) $\tau_{\rm D}$, as well as of the transmission time $\tau_{\rm T}$ and the reflection time $\tau_{\rm R}$ which were based on the time dependence of the probabilities $P_i$ $(i=1,2,3)$ of finding the particles in the region $x<-d$, $|x|<d$, $x>d$ respectively, the potential barrier lying between $x=-d$ and $x=d$.\\
According to these definitions
\be
\tau_{\rm D}=\int_{t_2}^{\infty} \, P_2(t) dt
\label{tau_D_1}
\ee
\be
\tau_{\rm T}=\int_{t_3}^{\infty} \, \left[1 - \frac{P_3(t)}{T}\right] dt
\label{tau_T_1}
\ee
\be
\tau_{\rm R}=\int_{t_1}^{\infty} \, \left[1 - \frac{P_1(t)}{R}\right] dt
\label{tau_R_1}
\ee
each of the times $\tau_{\rm D}$, $\tau_{\rm T}$ and $\tau_{\rm R}$ are weighted sums of the time intervals $dt$.
For $\tau_{\rm D}$ the weight is the probability of finding the particle within the barrier, for $\tau_{\rm T}$ and $\tau_{\rm R}$ the weight is the fraction of the total transmitted (reflected) packet which at time $t$ has not yet been transmitted (reflected), $T$ and $R$ being the probability of the particle being definitely transmissetted and reflected at $t \rightarrow \infty$ respectively. $\tau_{\rm D}$ is independent of the choice of the lower extremum $t_2$ in
eq. \ref{tau_D_1}, whereas $\tau_{\rm T}$ and $\tau_{\rm R}$ depend on the values of $t_1$ and $t_3$, which should be chosen as the time when the packet begins to interact with the barrier. In ref. \cite{bebr96} we fixed $t_1$ and $t_3$ as the time $\tau_{\epsilon}$ such that the contribution to the dwell time of the interval [0,$\tau_{\epsilon}$] is equal to that fraction $\epsilon$ of the total dwell time which can be regarded as the resolution of our procedure. We estimated this resolution as 0.01$\tau_{\rm D}$.

It is desirable, however, to remove also the slight arbitrariness implicit in the above choice, and to have a more objective definition of the times involved in the tunneling process. To this purpose, we resort to the well known device (see refs. \cite{baz}-\cite{ryba}-\cite{bu83}) of attaching to the particle a clock which runs only when the particle is in the barrier region. This clock can be provided by a magnetic moment which precesses around a constant magnetic field ${\bf B}$ confined within the barrier region. In order for this interaction not to affect substantially the barrier height, it is necessary to consider the limit of infinitesimal magnetic field, which will cause an infinitesimal precession. The measure of this precession, i.e. of the average of the spin of the particle in the region beyond the barrier (or the average over the reflected component) should yield the time the particle has spent in the barrier before being transmitted (or reflected). So far, the use of a physical clock for determining the tunneling time has been limited to the case of stationary solutions.

In order to work out this program, we have studied the evolution of a wave packet representing a particle with spin 1/2, polarized in the $x$ direction, initially located in front of the barrier, which travels in the $y$ direction towards the barrier. The average kinetic energy is lower than the barrier height. In the barrier region the particle interacts with a magnetic field ${\bf B}$ directed in the $z$ direction. As a consequence, the transmitted and reflected packets will be no longer polarized in the $x$ direction.

A magnetic moment ${\bf \mu}$ initially aligned in the $x$ direction which emerges at time $t$ from the barrier region after interacting a time $\tau(t)$ with a constant magnetic field has a $y$ component $\mu_y=-\mu \sin(\omega t)$ where (in units with $\hbar=1$) $\omega=\mu B$ is the Larmor frequency. In turn, the $x$ component is $\mu_x(t) = \mu \cos(\omega t)$. We extract $\tau(t)$ from the average, at time $t$, of the operators $\mu_x$ and $\mu_y$ over the transmitted packet, which are proportional to the average values over the transmitted packet of $S_x$ and $S_y$, which we denote as $\langle S_x(t) \rangle_3$ and $\langle S_y(t) \rangle_3$. We solve for $\tau(t)$ either of the equations
\be
\langle S_x(t) \rangle_3 = \frac{1}{2} \frac{\displaystyle{\int_0^t \, \cos[\omega \tau(x)] \frac{dP_3}{dx} dx}}{P_3(t)}
\label{S_x_clas_3}
\ee
\be
\langle S_y(t) \rangle_3 = -\frac{1}{2} \frac{\displaystyle{\int_0^t \, \sin[\omega \tau(x)] \frac{dP_3}{dx}} dx}{P_3(t)}
\label{S_y_clas_3}
\ee
The above equations are to be viewed as the definition of $\tau(t)$. The tunneling time is calculated as follows:
\be
\tau_{\rm T} = \frac{\displaystyle{\int_0^{\infty} \, \tau(t) \frac{dP_3}{dt}dt}}{P_3(\infty)}
\label{tau_T_2}
\ee
In other words, we weight every increment $dP_3$ of the probability of having the particle beyond the barrier with the time $\tau(t)$ that same fraction $dP_3$ of the transmitted packet has spent within the barrier, the time being measured through the spin precession. Throughout the calculation we take advantage of the infinitesimal value of $\omega$, so that we are content with the first non-vanishing term.

We notice that in principle eqs. \ref{S_x_clas_3} and \ref{S_y_clas_3} yield different definitions of $\tau(t)$, i.e. we have two different observables, $S_x$ and $S_y$, whose measurement allows an indirect measurement of $\tau(t)$. Thus, eq. \ref{tau_T_2} could yield two different results for the tunneling time $\tau_{\rm T}$. As a matter of fact, the results obtained starting from eq. \ref{S_x_clas_3} or eq. \ref{S_y_clas_3} are equal well within the accuracy of the calculation, although, as will be discussed in sec. 4, the values of $\tau(t)$ are not exactly equal (but they are practically equal starting from sufficiently early times). Moreover, the results are independent from the location of the packet at time $t=0$.

With analogous considerations we derive the reflection time $\tau_{\rm R}$, starting from the average of $S_x$ and $S_y$ in the region in front of the barrier:
\be
\langle S_x(t) \rangle_1 = \frac{1}{2} \frac{\displaystyle{\int_0^t \, \cos[\omega \tau(x)] \frac{dP_1}{dx} dx}}{P_1(t)}
\label{S_x_clas_1}
\ee
\be
\langle S_y(t) \rangle_1 = -\frac{1}{2} \frac{\displaystyle{\int_0^t \, \sin[\omega \tau(x)] \frac{dP_1}{dx} dx}}{P_1(t)}
\label{S_y_clas_1}
\ee
where the label 1 denotes the average over the region in front of the barrier. Also in this case the values obtained by choosing either of the values of $\tau(t)$ extracted from eqs. \ref{S_x_clas_1} and \ref{S_y_clas_1}
\be
\tau_{\rm R} = \frac{\displaystyle{\int_0^{\infty} \, \tau(t) \frac{dP_1}{dt} dt}}{P_1(\infty)}
\label{tau_R_2}
\ee
are very near and do not depend on the initial position of the wave packet.

In conclusion, the prescriptions embodied in eqs \ref{S_x_clas_3}-\ref{tau_T_2} (\ref{S_x_clas_1}-\ref{tau_R_2}) appear to give a rather objective measurement of the tunneling time $\tau_{\rm T}$ (of the reflection time $\tau_{\rm R}$). The resulting values are in agreement with those we found in ref. \cite{bebr96} and do not suffer of the ambiguity in the choice of the lower extremum in eqs. \ref{tau_T_1} and \ref{tau_R_1}, i.e. the time when the particle begins to interact with the potential.

In sec. 2 we present the problem, and in sec. 3 we derive the time $\tau(t)$ and the consequent values for $\tau_{\rm R}$ and $\tau_{\rm T}$. We comment our results in sec. 4, while sec. 5 is devoted to the conclusions.

\section{Definition of the problem}

We consider a wave packet describing a particle of mass $m$, moving in the $y$ direction, where a potential barrier lies between $y=-d$ and $y=d$:
\be
V(y) = V_0 = \frac{k_0^2}{2m} \quad|y| < d
\label{potential}
\ee
A homogeneous magnetic field ${\bf B}$ directed in the $z$ direction is confined in the barrier region. The particle has a magnetic moment ${\bf \mu} =g{\bf S}$ which in the barrier region is coupled to the magnetic field by the interaction ${\bf \mu} \cdot {\bf B} = g B S_z$. Consequently, the hamiltonian of the particle is (we use units with $\hbar=1$):
\be
H=
\left \{ \begin{array}{ll}
\displaystyle{\frac{p^2}{2m} \qquad} & ~(|y|>d) \\ 
\\
\displaystyle{\frac{p^2}{2m}+V_0 \qquad} & ~(|y|<d)
\end{array} \right.
\label{hamiltonian}
\ee
The initial state of the particle is described by the wave function
\be
\psi(y,0) = \phi(y) \chi
\label{psi_0}
\ee
where $\chi$ is the spinor $(1/\sqrt{2})(1,1)$, such that $S_x \chi = \chi$ and $\phi(y)$ is a Gaussian packet located in front of the barrier and peaked at $y=y_0$. We choose for $\phi(y)$ the same wave function whose evolution was studied in ref. \cite{bebr96},

\be
\phi(y) = \int \, a(k) \psi_k(y) dk
\label{phi_0}
\ee
where $\psi_k(y)$ are the eigenfunctions of the hamiltonian \ref{hamiltonian} with $\omega=0$ and
\be
a(k) = \left[\frac{2 \delta^2}{4 \pi^3}\right]^{1/4} e^{-(k-k_{\rm av})^2} e^{-i k y_0}
\label{ak}
\ee
The parameters involved in the problem are chosen as follows:
\be
m = 1,\,\,d = \sqrt{2},\,\,k_{\rm av} = 9.9,\,\,d = 2,\,\,k_0 = 10 ,\,\,y_0=-15
\label{parameters}
\ee

In order to study the time evolution of the wave function \ref{psi_0} under the action of the hamiltonian \ref{hamiltonian} we must expand \ref{psi_0} in the basis of the eigenfunctions $\psi_{k,+}(y)$ and $\psi_{k,-}(y)$ with $S_z=1/2$ and $S_z=-1/2$ respectively. These functions are the eigenfunctions of the square barrier hamiltonian, where the barrier height is $V_0-\omega/2$ and $V_0+\omega/2$ respectively. We note however that the initial wave function is located in the region $y<-d$, where the dependence of $\psi_{k,+}(y)$ and $\psi_{k,-}(y)$ on $S_z$ is only through the reflection coefficient $A(k)$. The contribution of this part of the wave function to the coefficients $a_+(k)=\int\, \psi^*_{k,+}(y)\phi(y) dy$ and $a_-(k)=\int\, \psi^*_{k,-}(y)\phi(y) dy$ is negligible, due to the Gaussian shape of \ref{ak}.
Consequently, we can safely assume the coefficients $a_+(k)$ and $a_-(k)$ to be the same, equal to $a(k)$ as given in eq. \ref{ak}.

The time evolution of the wave function \ref{psi_0} is given by
$$\int \, a(k) \psi_{k,+}(y) e^{-ik^2t/2m}dk$$
that is
\be
\psi(y,t) = \frac{1}{\sqrt{2}}\int \, a(k) \psi_{k}(y) e^{-ik^2t/2m}dk
\label{psi}
\ee

\section{Evaluation of the tunneling time and the reflection time}

We derive the tunneling and reflection times according to the program outlined in the Introduction.
We focus our attention on the calculation of the tunneling time, the case of the reflection time being quite similar.

Let $P_3(t)$ be the probability of finding the particle beyond the barrier
($y > d$) at time $t$ ($\Theta$ is the Heavyside function):
\be
P_3(t) = (\psi(y,t) , \Theta(y-d) \psi(y,t))
\label{P_3}
\ee
We want to associate to the increment $dP_3 = dP_3/dt\,dt$ of this probability the time $\tau(t)$ which this fraction of the transmitted packet has spent in the barrier region.
The tunneling time for the packet will be calculated according to the equation \ref{tau_T_2}
$$\tau_{\rm T} = \frac{\displaystyle{\int_0^{\infty} \, \tau(t) \frac{dP_3}{dt} dt}}{P_3(\infty)}$$
The time $\tau(t)$ will be extracted from the average of the spin component $S_x$ or $S_y$ at time $t$ in the region beyond the barrier:
\be
\langle S_x(t) \rangle_3 = \frac{(\psi(y,t) , \Theta(y-d)S_x \psi(y,t))}{P_3(t)}
\label{S_x_3}
\ee
\be
\langle S_y(t) \rangle_3 = \frac{(\psi(y,t) , \Theta(y-d)S_y \psi(y,t))}{P_3(t)}
\label{S_y_3}
\ee
The key for deriving $\tau(t)$ is the spin precession which occurs when the particle is in the barrier region where a constant magnetic field aligned with the $z$ axis is located.
For a magnetic moment ${\bf \mu}$ aligned with the $x$ axis at time $t=0$ and interacting with the field, the time dependence of its $x$- and $y$-component is given by
\be
\mu_x = \mu \cos(\omega t) \qquad {\rm and} \qquad \mu_y = -\mu \sin(\omega t)
\label{mu_i}
\ee
where $\omega = \mu B$. The same time dependence holds for the $x$ and $y$ components of the spin angular momentum:
\be
S_x = \frac{1}{2} \cos(\omega t) \qquad {\rm and} \qquad S_y = - \frac{1}{2} \sin(\omega t)
\label{S_i}
\ee
If the fraction $dP_3$ of the packet emerging beyond the barrier in the time interval dt at time $t$ has spent a time $\tau(t)$ in the barrier region, we assign to this fraction a contribution to $\langle S_x(t) \rangle_3$ and $\langle S_y(t) \rangle_3$ equal to
\be
dS_x = \frac{1}{2} \frac{dP_3}{dt} \cos[\omega \tau(t)] dt
\label{dS_x_3}
\ee
\be
dS_y = - \frac{1}{2} \frac{dP_3}{dt} \sin[\omega \tau(t)] dt
\label{dS_y_3}
\ee
Consequently, the values of $\langle S_x(t) \rangle_3$ and $\langle S_y(t) \rangle_3$ will be given as
\be
\langle S_x(t) \rangle_3 = \frac{1}{2}\frac{\displaystyle{\int_0^t \frac{dP_3}{dp} \cos[\omega \tau(p)] dp}}{P_3(t)}
\label{av_S_x_clas_3}
\ee
\be
\langle S_y(t) \rangle_3 = -\frac{1}{2}\frac{\displaystyle{\int_0^t \frac{dP_3}{dp} \sin[\omega \tau(p)] dp}}{P_3(t)}
\label{av_S_y_clas_3}
\ee
We have to extract $\tau(t)$ from either of eqs. \ref{av_S_x_clas_3} and \ref{av_S_y_clas_3}. This can be done by multiplying both sides of the equations by $P_3(t)$ and taking the derivative with respect to $t$. It is useful to observe, however, that we need to consider the limit when $\omega$ tends to zero, since we are interested in the tunneling time beyond a barrier of heigth $V_0$, and a non-infinitesimal magnetic field would alter the barrier heigth.

If we write
\be
\langle S_x(t) \rangle_3 = \frac{N_x(t)}{P_3(t)} \qquad {\rm and} \qquad \langle S_y(t) \rangle_3 = \frac{N_y(t)}{P_3(t)}
\label{S_notations}
\ee
we get:
\be
\frac{dN_x(t)}{dt} = \frac{1}{2} \frac{dP_3(t)}{dt} \cos[\omega \tau(t)]
\label{dN_x_3_dt}
\ee
\be
\frac{dN_y(t)}{dt} = -\frac{1}{2} \frac{dP_3(t)}{dt} \sin[\omega \tau(t)]
\label{dN_y_3_dt}
\ee
In eq. \ref{dN_x_3_dt} we approximate $\cos(\omega t) \simeq 1 - \frac{1}{2} \omega^2 t^2$, and in eq. \ref{dN_y_3_dt} we put $\sin(\omega t) \simeq \omega t$. We note also that $N_x$ and $P_3$ are even in $\omega$, whereas $N_y$ is odd in $\omega$. Moreover, at order 0 in $\omega$ we have $N_x=\frac{1}{2} P_3$, $N_y=0$. By putting
$$N_x=N_x^{(0)} + \omega^2 N_x^{(2)}$$
$$P_3=N_x^{(0)} + \omega^2 P_3^{(2)}$$
$$N_y=\omega N_y^{(1)}$$
from eqs. \ref{dN_x_3_dt} and \ref{dN_y_3_dt} we get
\be
\frac{dN_x^{(2)}(t)}{dt}=\frac{1}{2} \left[ \frac{dP_3^{(2)}(t)}{dt}-\frac{1}{2} \tau(t)^2 \frac{dN_x^{(0)}(t)}{dt} \right]
\label{x_der}
\ee
\be
\frac{dN_y^{(1)}(t)}{dt} = -\frac{1}{2} \frac{dN_x^{(0)}(t)}{dt} \tau(t)
\label{y_der}
\ee
According to the above equations, we have two different possible prescriptions to derive the time $\tau(t)$ which must be inserted into eq. \ref{tau_T_2}. We denote by $\tau_{\rm T,x}$ the result obtained from eq. \ref{tau_T_2} using eq. \ref{x_der}, which yields for $\tau(t)$
\be
\tau(t) = \sqrt{2\frac{\left(\displaystyle{\frac{dP_3^{(2)}}{dt} -2 \frac{dN_x^{(2)}}{dt}}\right)}{\displaystyle{\frac{dN_x^{(0)}}{dt}}}}
\label{x_tau}
\ee
and by $\tau_{\rm T,y}$ the result for $\tau(t)$ obtained from eq. \ref{y_der}:
\be
\tau(t) = -2 \frac{\displaystyle{\frac{dN_y^{(1)}}{dt}}}{\displaystyle{\frac{dN_x^{(0)}}{dt}}}
\label{y_tau}
\ee
The calculation of $\tau_{\rm T,x}$ requires the calculation of the integral
\be
\tau_{\rm T,x} = \frac{1}{P_3(\infty)} \int_0^{\infty} \sqrt{2\frac{dN_x^{(0)}}{dt}\left(\frac{dP_3^{(2)}}{dt} - 2\frac{dN_x^{(2)}}{dt} \right)} dt
\label{x_tau_T}
\ee
whereas, according to eq. \ref{tau_T_2} and eq. \ref{y_der}, the value of $\tau_{\rm T,y}$ depends only on the values at $t=\infty$ of $N_y^{(1)}$ and $P_3$, and is given immediately by
\be
\tau_{\rm T,y} = -\frac{2}{P_3(\infty)} \int_0^{\infty} \frac{dN_y^{(1)}}{dt} dt = - 2 \frac{N_y^{(1)}(\infty)}{P_3(\infty)}
\label{y_tau_T}
\ee
Performing the calculations, we find:
\be
\tau_{\rm T,x} = 3.53 \qquad \qquad \tau_{\rm T,y} = 3.52
\label{tau_T_2_num}
\ee
In order to find $\tau_{\rm R}$ we repeat the above calculations starting from the values of
$$\langle S_x(t) \rangle_1 = \frac{N_x(t)}{P_1(t)} \qquad {\rm and} \qquad \langle S_y(t) \rangle_1= \frac{N_y(t)}{P_1(t)}$$
 where now
$$N_x(t) = (\psi(y,t),\Theta(-y-d) S_x \psi(y,t)) \qquad {\rm and} \qquad N_y(t) = (\psi(y,t),\Theta(-y-d)S_y\psi(y,t))$$
These allow the extraction of $\tau(t)$ according to formulae analogous to eqs. \ref{x_tau} and \ref{y_tau}, where of course $P_3$ is substituted with $P_1$. The reflection times $\tau_{\rm R,x}$ and $\tau_{\rm R,x}$ are given by the equations
\be
\tau_{\rm R,x} = \frac{1}{P_1(\infty)} \int_0^{\infty} \sqrt{2\frac{dN_x^{(0)}}{dt}\left(\frac{dP_1^{(2)}}{dt} - 2\frac{dN_x^{(2)}}{dt} \right)} dt
\label{x_tau_R}
\ee
\be
\tau_{\rm R,y} = -\frac{2}{P_1(\infty)} \int_0^{\infty} \frac{dN_y^{(1)}}{dt} dt = - 2 \frac{N_y^{(1)}(\infty)}{P_1(\infty)}
\label{y_tau_R}
\ee
The results are as follows:
\be
\tau_{\rm R,x} = 0.52 \qquad \qquad \tau_{\rm R,y} =0.56
\label{tau_R_2_num}
\ee
Finally, we observe that, while eqs. \ref{tau_T_2_num} and \ref{tau_R_2_num} directly connect the average value at $t=\infty$ of $S_y$ in the region in front or beyond the barrier with $\tau_{\rm R,x}$ or $\tau_{\rm T,x}$, this connection does not hold for the $x$-component, i.e. we do not have the relation, which holds in the stationary problem \cite{bu83},
\be
\langle S_x \rangle = \frac{1}{2}\left(1 - \frac{1}{2} \omega^2 \tau_x^2\right)
\label{buttiker_cond}
\ee
As a consequence, there is no use in introducing $\tau_z$ as $\tau_z = 2 \langle S_z \rangle_/\omega$, since the condition $\tau_z^2 = \tau_x^2 + \tau_y^2$, which is derived from the identity $\langle S_x \rangle^2 + \langle S_y \rangle^2+ \langle S_z \rangle^2=1/4$, no longer holds, due to the failure of eq. \ref{buttiker_cond}.

\section{Comments}

The results obtained in the previous section for both the tunneling and the reflection time using $\tau_x(t)$ or $\tau_y(t)$ are near to each other, and are remarkably similar to the values found in ref. \cite{bebr96}, where we found $\tau_{\rm T} = 3.39$, $\tau_{\rm R}=0.55$. This supports the reliability of the definition of the tunneling and reflection times proposed therein and the conclusions about the impossibility of extending the notion of tunneling (reflection) time proposed for stationary problems by several authors to the case of wave packets. For a comparative discussion of some results of this kind we defer the reader to ref. \cite{bebr96}.

Some comments are in order with regard to the dependence of the tunneling and reflection time on the initial position of the packet and on the different ways to evaluate this time (i.e. the choice between $\tau_{\rm T,x}$ and $\tau_{\rm T,y}$, or between $\tau_{\rm R,x}$ and $\tau_{\rm R,y}$).

Concerning the first point, it is easy to see that the integral in eq. \ref{y_tau_T} can be written as
\be
\tau_{\rm T,y} = \frac{\displaystyle{\int \, |a(k)|^2 {\rm Im}\left[D(k)^* \frac{\partial D(k)}{\partial \omega}\right] dk}}{\displaystyle{\int \, |a(k)|^2|D(k)|^2}}
\label{y_tau_T_asy}
\ee
where $D(k)$ is the transmission coefficient and the derivative with respect to $\omega$ is taken at $\omega=0$. No dependence on the value $y_0$ where the packet is peaked at $t=0$ is left.

The comparison between $\tau_{\rm T,x}$ and $\tau_{\rm T,y}$ ($\tau_{\rm R,x}$ and $\tau_{\rm R,y}$) is more delicate. We will focus our discussion on the tunneling time, the considerations for the reflection time being quite similar. Eqs. \ref{dN_x_3_dt} and \ref{dN_y_3_dt} show that the condition for the identity of the values of $\tau(t)$ derived from them is
\be
\left[\frac{dN_x(t)}{dt}\right]^2 + \left[\frac{dN_y(t)}{dt}\right]^2 = \frac{1}{4} \left[\frac{dP_3(t)}{dt}\right]^2
\label{identity_cond}
\ee
This condition is not verified for any value of $t$. However, the ratio 
\be
R=\frac{\displaystyle{4 \left[\frac{dN_x(t)}{dt}\right]^2 + 4 \left[\frac{dN_y(t)}{dt}\right]^2 - \left[\frac{dP_3(t)}{dt}\right]^2}}{\displaystyle{\left[\frac{dP_3(t)}{dt}\right]^2}}
\label{ratio}
\ee
is less than 0.1 starting from $t \simeq 2$ (less than 0.01 starting from $t \simeq 2.7$), where the main contribution to the integral \ref{tau_T_2} comes from. This can be easily understood. With the parameters of the problem, the packet travels with a velocity approximately equal to $k_{\rm av}/m \simeq 10$. Given that the initial peak is located at $y_0=-15$, up to $t \simeq 1.5$ the fraction of the packet in the barrier region is practically zero.

As to the reasons why the LHS of eq. \ref{identity_cond} is very nearly equal to the RHS, while not being exactly equal, they can be traced to the following observations. Each term in the numerator of eq. \ref{ratio} can be written in terms of the corresponding flux calculated at $y=d$. If $\psi_+$ ($\psi_-$) and $\psi_+^{\prime}$ ($\psi_-^{\prime}$) are the values in $y=d$ of the component with $S_z=1/2$ ($S_z=-1/2$) of the wave function and its derivative with respect to $y$, the numerator $N$ in eq. \ref{ratio} reads
\be
N = \left|\psi_+^{\prime \, *}\psi_- - \psi_+^* \psi_-^{\prime}\right|^2 - \left|j_+ + j_-\right|^2
\label{numerator}
\ee
$j_+$ and $j_-$ being the currents due to $\psi_+$ and $\psi_-$, calculated in $y=d$. If the
function $\psi_+$ ($\psi_-$) is expressed in polar form, $\psi_+ = \rho_+ e^{i\theta_+}$, ($\psi_- = \rho_- e^{i\theta_-}$), we have $j_+ = \rho_+^2\theta_+^{\prime}$ ($j_- = \rho_-^2\theta_-^{\prime}$) and $N$ can be written as follows:
\be
N = (\rho_-\rho_+^{\prime} - \rho_+ \rho_-^{\prime})^2 - (\rho_+^2\theta_+^{\prime \, 2} - \rho_-^2\theta_-^{\prime \, 2})(\rho_+^2 - \rho_-^2)
\label{polar_N}
\ee
At order 0 in $\omega$ $N$ vanishes, whereas the second order contribution in $\omega$ is due only to the first order variation, i.e. it can be obtained by writing $\rho_+ = \rho - \omega d\rho$, $\rho_- = \rho + \omega d\rho$ and the likes for the other terms. As a consequence eq. \ref{numerator} can be written as follows:
\be
R =\omega^2 \frac{4 \left(\rho^{\prime}d\rho - \rho d\rho^{\prime}\right)^2
-16\rho^2\theta^{\prime}d\rho(\theta^{\prime}d\rho + \rho d\theta^{\prime})}{\rho^4 \theta^{\prime \, 2}}
\label{ratio_1st}
\ee
On the other hand, using eq. \ref{dN_x_3_dt} and \ref{dN_y_3_dt} and denoting by $\tau_x$ the value of $\tau(t)$ in eq. \ref{dN_x_3_dt}
and by $\tau_y$ the value of $t$ in eq. \ref{dN_y_3_dt}, at second order in $\omega$ $R$ can be written as
\be
R = \omega^2(\tau_y^2 - \tau_x^2) = 2\omega^2\tau_{\rm av} \Delta \tau \label{ratio_2nd}
\ee
$\Delta \tau$ being $\tau_y - \tau_x$ and $\tau_{\rm av}$ the average of $\tau_x$ and $\tau_y$. As a consequence
\be
\frac{\Delta \tau}{\tau_{\rm av}} = \frac{2(\rho^{\prime}d\rho - \rho d\rho^{\prime})^2 -8\rho^2\theta^{\prime}d \rho(\theta^{\prime} d \rho + \rho d\theta^{\prime})}{\rho^4\theta^{\prime \, 2}\tau_{\rm av}^2}
\label{last_cond}
\ee
Since $d\rho/\rho$ and the analogous terms for $\rho^{\prime}$ appearing in eq. \ref{last_cond} are much smaller than 1, we have $\Delta \tau/ \tau_{\rm av} \ll 1$, and consequently $\tau_{\rm T,x}$ = $\tau_{\rm T,y}$ .

\section{Conclusions}

We have studied the process of a wave packet tunneling beyond a potential barrier which has a clock running only in the barrier region. This is achieved by the well known device, so far envisaged only for stationary problems, of considering the particle endowed with a magnetic moment coupled to a magnetic field confined in the barrier region. The time is measured by means of the precession of the magnetic moment around the field direction. 

The particle is initially polarised in a direction orthogonal to the direction of motion and that of the field. So, for any fraction of the packet emerging beyond the barrier (or reflected by it), the value of each of the two components of the magnetic moment beyond the barrier (or in front of it) orthogonal to the field can yield a measurement of the time spent in the barrier region. We find that the results obtained using any of these components are quite similar, within the accuracy of the calculation. The results for the tunneling and reflection times obtained by means of this method are in good agreement with the results obtained for the same packet, based on the consideration of the time behaviour of the probabilities of having the particle in front or within or beyond the barrier \cite{bebr96}.

The tunneling and reflection times obtained in this way are independent of the position at time $t=0$ of the packet, provided it is located sufficiently far from the barrier edge. Thus, the definition of the tunneling or reflection time by means of a physical clock looks a promising approach when one wants to go beyond the monochromatic approximation.

\end{document}